\documentclass[pdflatex,sn-mathphys-num]{sn-jnl}
\usepackage{lineno}

\usepackage{graphicx}%
\usepackage{multirow}%
\usepackage{amsmath,amssymb,amsfonts}%
\usepackage{amsthm}%
\usepackage{mathrsfs}%
\usepackage[title]{appendix}%
\usepackage{xcolor}%
\usepackage{textcomp}%
\usepackage{manyfoot}%
\usepackage{booktabs}%
\usepackage{algorithm}%
\usepackage{algorithmicx}%
\usepackage{algpseudocode}%
\usepackage{listings}%
\usepackage[T1]{fontenc}

\usepackage{aas_macros}

\theoremstyle{thmstyleone}%
\theoremstyle{thmstyletwo}%

\theoremstyle{thmstylethree}%

\definecolor{myblue}{rgb}{0.13, 0.13, 0.8}

\begin{document}

\title[Overmassive black holes in the early Universe]{Overmassive black holes and little red dots naturally form in simulations}

\author*[1]{\fnm{Sunmyon} \sur{Chon}}\email{sunmyon@MPA-Garching.MPG.DE}

\author[2,3]{\fnm{Shingo} \sur{Hirano}}\email{shingo-hirano@kanagawa-u.ac.jp}

\author[4]{\fnm{Tomoaki} \sur{Ishiyama}}\email{ishiyama@chiba-u.jp}

\author[1]{\fnm{Seok-Jun} \sur{Chang}}\email{sjchang@MPA-Garching.MPG.DE}

\author[1]{\fnm{Volker} \sur{Springel}}\email{vspringel@MPA-Garching.MPG.DE}

\affil*[1]{\orgname{Max-Planck-Institut f\"ur Astrophysik}, \orgaddress{\street{Karl-Schwarzschild-Str. 1}, \city{Garching}, \postcode{D-85741}, \state{Garching bei M\"unchen}, \country{Germany}}}

\affil[2]{\orgdiv{Department of Applied Physics, Faculty of Engineering}, \orgname{Kanagawa University}, \orgaddress{\street{3-27-1 Rokukakubashi, Kanagawa, Yokohama}, \city{Kanagawa}, \postcode{221-0802}, \country{Japan}}}

\affil[3]{\orgdiv{Department of Astronomy, School of Science}, \orgname{University of Tokyo}, \orgaddress{\street{7-3-1 Hongo, Bunkyo}, \city{Tokyo}, \postcode{113-0033}, \country{Japan}}}

\affil[4]{\orgdiv{Digital Transformation Enhancement Council}, \orgname{Chiba University}, \orgaddress{\street{1-33, Yayoi-cho, Inage-ku}, \city{Chiba}, \postcode{263-8522}, \country{Japan}}}


\abstract{
The origin of supermassive black holes (SMBHs) remains a long-standing problem in astrophysics. 
Recent JWST observations reveal an unexpectedly abundant population of overmassive black holes at $z>4$--$6$, where the BH masses lie far above local scaling relations and are not reproduced by current cosmological models \citep{Harikane+2023, Ubler+2023, Maiolino+2024, Maiolino+2024Nat, Taylor+2025}. 
How such overmassive black holes form and rapidly grow within young galaxies has remained unclear.
Here we present fully cosmological radiation-hydrodynamic simulations that self-consistently follow the birth, early growth, and emergent observable signatures of SMBHs in proto-cluster environments. 
We find that heavy seeds of order $10^{6}\,M_\odot$ naturally form, 
exceeding typical theoretical expectations by an order of magnitude. 
These seeds rapidly develop dense, optically thick disks whose strong electron scattering produces broad H$\alpha$ emission comparable to that seen in little red dots (LRDs) \citep{Matthee+2024, Hviding+2025, Rusakov+2025, Naidu+2025, deGraaff+2025}. 
Sustained super-Eddington accretion then drives fast growth to $\sim3\times10^{7}\,M_\odot$ by $z\simeq 8$.
This is the first demonstration that unifies LRDs to a short-lived, enshrouded phase of heavy-seed formation, which naturally evolve into the overmassive quasars detected by JWST and ultimately the progenitors of today’s SMBHs.
}


\maketitle

Supermassive black holes (SMBHs) are known to exist less than a billion years after the Big Bang, yet how they were seeded and grew remains unclear. Recent \textit{James Webb Space Telescope} (JWST) observations have revealed compact, red sources at $z>4$–$6$, the so-called little red dots (LRDs), whose inferred black-hole (BH) masses exceed local scaling relations \citep{Harikane+2023, Ubler+2023, Maiolino+2024, Maiolino+2024Nat, Taylor+2025}. Their spectra point to rapid BH growth in dense, obscured environments during early galaxy assembly \citep{Matthee+2024, Hviding+2025, Rusakov+2025, Naidu+2025, deGraaff+2025}, consistent with recent theoretical predictions \citep{Lupi+2024, Trinca+2024, Jeon+2025}. However, proposed pathways such as direct-collapse BHs or Population III remnants rely on idealized conditions and have not been followed self-consistently in cosmological simulations \citep{Matthee+2025, Bulichi+2025}. Here we show that heavy BH seeds naturally form in overdense proto-cluster regions exposed to intense far-ultraviolet (FUV) radiation, where the collapse of supermassive stars produces $\sim10^{6}\,M_\odot$ seeds that undergo brief super-Eddington growth. These systems reproduce the Balmer features and red continua seen in LRDs and rapidly grow into overmassive BHs by $z\sim8$. Our results provide a unified pathway linking the birth of massive seeds, their short-lived obscured growth phases, and the overmassive BHs discovered by JWST, offering a cosmological explanation for their abundance and properties.

\vspace{0.5em}
\noindent{\textbf{Rapid emergence of overmassive BHs}}

\begin{figure}[h]
\centering
\includegraphics[width=0.9\textwidth]{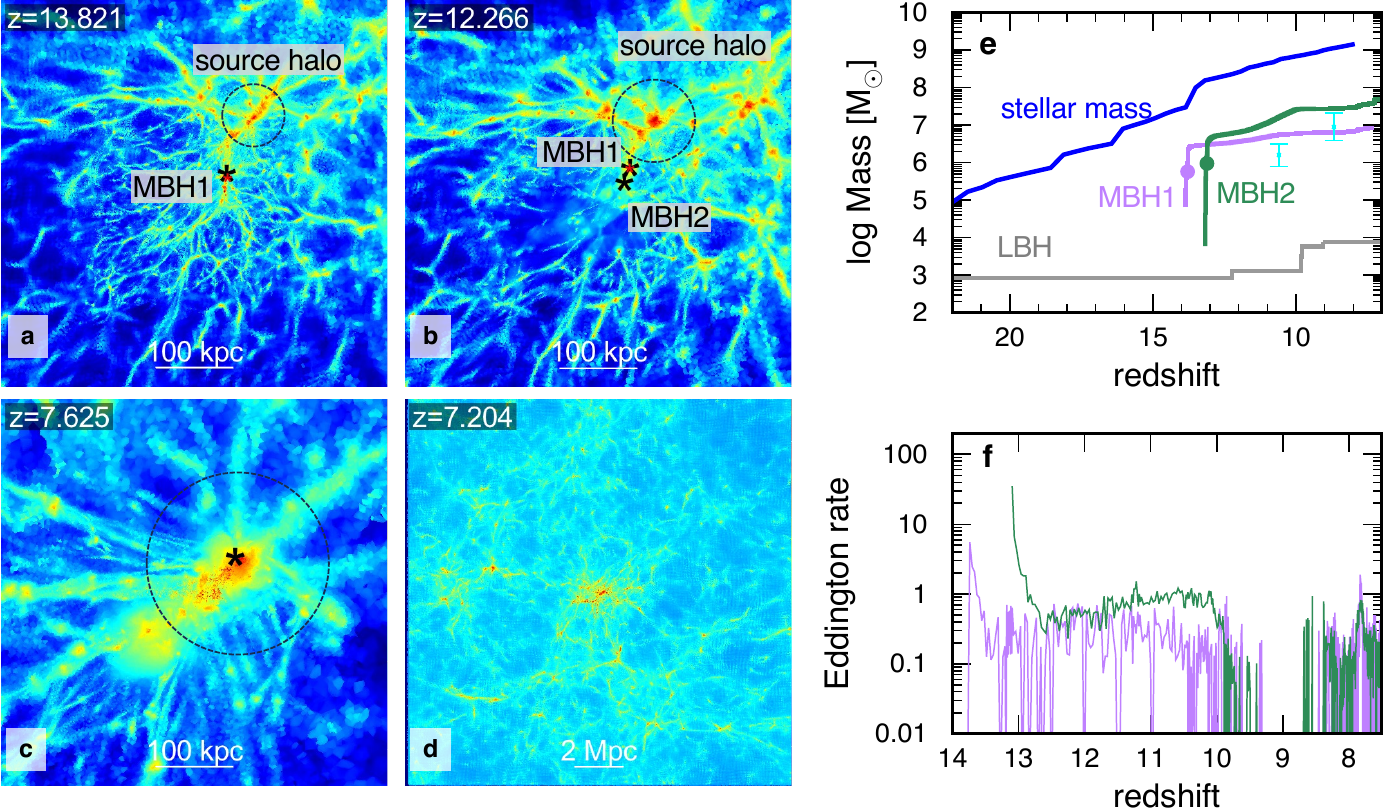}
\caption{Formation and evolution of massive seed BHs in the cosmological radiation-hydrodynamic simulation.  
(a–c) Gas density distributions around the target halos at representative epochs. Black asterisks indicate heavy-seed BHs formed via the collapse of supermassive stars, while red asterisks denote light-seed BHs originating from Population~III remnants. The latter almost overlaps with the position of MBH1. 
Panels (a) and (b) correspond to the epochs when the first (MBH1) and second (MBH2) heavy seeds form, respectively. In panel (c), both MBH1 and MBH2 have migrated into a nearby massive halo and subsequently merged. Dashed circles mark the neighboring source halos that provide the strong FUV radiation required for heavy-seed formation. The radii of the circles indicate their virial radii at each snapshot.  
(d) Large-scale gas density distribution across the entire simulated volume (16 $h^{-1}$ Mpc on a side), centered on the target halo, shown at the final simulation snapshot.
(e) Redshift evolution of the BH masses and the stellar mass of the source galaxy. The gray line shows the growth of a light-seed BH formed at $z\simeq22$, whereas the purple and green lines correspond to the two heavy seeds (MBH1 and MBH2). Circles mark the moments when the supermassive stars collapse into BHs. The blue line denotes the total stellar mass of the host galaxy. Cyan points with error bars show observed high-$z$ SMBHs \citep{Larson+2023, Maiolino+2024Nat}.
(f) Eddington ratios of the two heavy seeds as a function of redshift. Both MBH1 (purple) and MBH2 (green) undergo short phases of super-Eddington accretion.
}\label{fig1:snapshots}
\end{figure}

Our cosmological radiation–hydrodynamic simulations naturally produce massive BH seeds that grow into overmassive ($\gtrsim10^{7}\,M_\odot$) BHs by $z\sim10$.
This rapid growth occurs in overdense proto-cluster regions exposed to intense FUV radiation from nearby star-forming galaxies, where the radiation suppresses early star formation.
We follow the collapse of one such halo using three-dimensional radiation-hydrodynamic simulations performed with the moving-mesh code \textsc{\small AREPO} \citep{Springel2010} (see Methods). The halo—identified in \citet{Ishiyama+2025} as a promising heavy-seed site—is located $\sim10$ kpc from a luminous neighbour that provides the strong FUV flux needed to form heavy seeds (Extended Data Fig.~\ref{figex2:J21_evo}). As a result, the halo accumulates a large gas reservoir before collapsing at $z\simeq14$ (Fig.~\ref{fig1:snapshots}a,b). Once collapse sets in, the central protostars grow rapidly, reaching $5$–$9\times10^{5}\,M_\odot$, well above the canonical $\sim10^{5}\,M_\odot$ predicted by standard direct-collapse models \citep{Bromm+2003, Chon+2018, Regan+2020, Latif+2022, Kiyuna+2024}.
The unusually deep potential well of the host halo (virial temperature $\sim4\times10^{4}$ K; Extended Data Fig.~\ref{figex1:Mhalo_evo}) enables ionized gas to remain gravitationally bound and sustain the high accretion rates needed to form such massive seeds. Multiple seed-forming sites emerge within the same overdense environment, indicating that heavy-seed formation can naturally occur in proto-cluster regions.

These massive seeds do not remain static; each heavy seed subsequently experiences a brief but intense phase of super-Eddington accretion (Fig.~\ref{fig1:snapshots}f).
Immediately after collapse, a dense and optically thick envelope develops around the nascent BH, efficiently feeding the accretion disk.
Radiation trapping becomes important: photons are advected inward faster than they can diffuse outward, enabling accretion rates several to a few tens of times the Eddington limit for less than a million years \citep{Begelman1978, Ohsuga+2005, Sadowski+2016}.
The inflowing gas is supplied from the circum-BH disk, inside which the gas loses angular momentum through gravitational torques.
This sustains rapid growth until the reservoir is exhausted, where the BHs reach several $\times10^{6}~M_\odot$ shortly after formation.
The accretion rate then gradually decreases to $0.1$–$1$ times the Eddington limit as the gas supply dwindles, allowing the BHs to grow to $\gtrsim10^{7} ~M_\odot$ by redshift $z\simeq10$.

At $z\simeq10$, the accretion rate sharply declines to below 1\% of the Eddington limit.
As the BHs pass through the central region of the massive neighboring halo, their envelope gas is stripped by the surrounding hot medium \citep{Chon+2021}.
Dynamical friction subsequently brings them to the galaxy center, where they merge with the host system.
By redshift $z\simeq8$, the BHs have settled in the central kiloparsec of a galaxy with stellar mass $\sim10^{9}~M_\odot$, establishing a BH-to-stellar-mass ratio of $\sim1\%$, an order of magnitude above the local relation.
After settling, the BHs resume sub-Eddington accretion, maintaining slow but steady growth.
For a brief period, the accretion rate reaches the Eddington limit, during which the inferred BH masses and accretion luminosities are consistent with those of observed high-redshift AGN \citep{Larson+2023}.
Our simulation yields a lower limit of a comoving number density of $\sim10^{-4}~\mathrm{Mpc^{-3}}$ for massive BH–hosting galaxies, consistent with that required to explain the observed LRDs \citep{Harikane+2023, Matthee+2024, Maiolino+2024, Taylor+2025}.

To demonstrate that such rapid growth is unique to heavy seeds, we also follow the evolution of BHs seeded by Population III remnants ($\sim800~M_\odot$) formed earlier than the heavy seeds at redshift $z\simeq22$, hereafter referred to as a light-seed black hole (LBH).
Their growth remains highly inefficient. The initial gas supply is strongly suppressed by radiative feedback from the Population III stars, resulting in final masses orders of magnitude smaller than those of the massive seeds (gray line in Fig.~\ref{fig1:snapshots}e).
The gas in the host halo quickly photo-evaporates during seed formation because its shallow gravitational potential cannot confine the fuel required for accretion.
This contrast highlights the necessity of heavy-seed formation to account for the overmassive BHs already present at $z>7$.

\begin{figure}[h]
\centering
\includegraphics[width=0.8\textwidth]{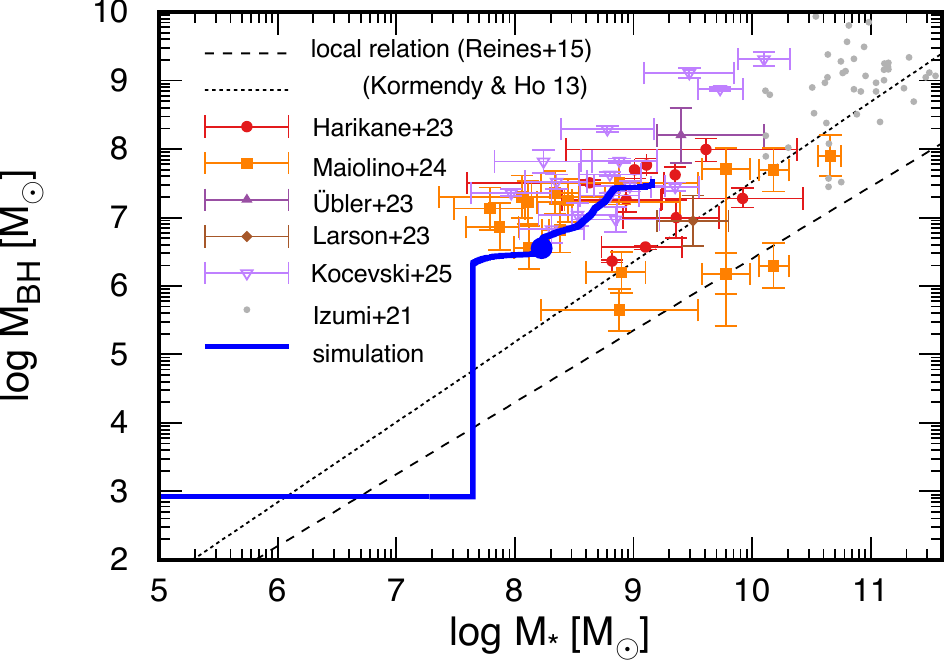}
\caption{
Time evolution of the most massive BH and the stellar mass of the nearby source galaxy.
The solid line shows the simulated evolution of the BH mass ($M_{\mathrm{BH}}$) as a function of the host-galaxy stellar mass ($M_\ast$).
Observational estimates for LRDs from recent JWST studies \citep{Harikane+2023, Ubler+2023, Larson+2023, Maiolino+2024, Kocevski+2025} and for quasars \citep{Izumi+2021} are overplotted for comparison (note that \citealt{Izumi+2021} uses dynamical masses rather than stellar masses).
The dotted and dashed lines indicate the local $M_{\mathrm{BH}}$–$M_\ast$ relations derived from nearby galaxies \citep{Kormendy+2013, Reines+2015}.
The blue point marks the time when the BH enters the virial radius of the host halo.
Our model naturally reproduces the overmassive BHs observed in LRDs, with $M_{\mathrm{BH}}/M_\ast$ ratios up to an order of magnitude above the local relation.
}\label{fig2:MBH_Mstar}
\end{figure}

Fig.~\ref{fig2:MBH_Mstar} shows the time evolution of the most massive BH and the stellar mass of the source galaxy.
Points with error bars denote the observed BH and stellar masses of the LRDs, while the blue line traces the simulated relation.
Before the formation of the massive seed, the BH mass lies far below the observed values.
After the emergence of the heavy seed and its brief phase of super-Eddington growth, the simulated BH mass becomes comparable to the observations, reaching the BH-to-stellar mass ratio above 0.01.
Subsequently, the BH and stellar masses co-evolve, while the BH mass remains well above the local scaling relation indicated by the dashed line.

\vspace{0.5em}
\noindent{\textbf{Proto-supermassive stellar system}}

\begin{figure}[h]
\centering
\includegraphics[width=0.9\textwidth]{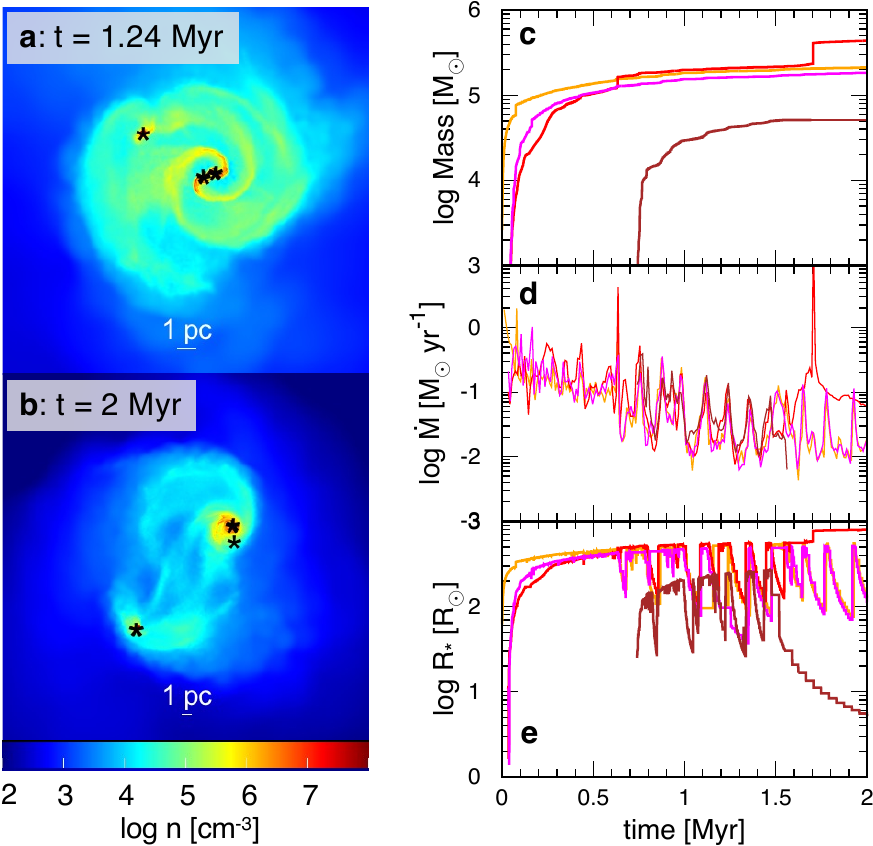}
\caption{Formation and evolution of protostars that grow into supermassive stars, which subsequently collapse into MBH1.  
(a, b) Projected gas density distributions at $t = 1.24~\mathrm{Myr}$ (panel a) and $t = 2.0~\mathrm{Myr}$ (panel b), where time is measured from the formation of the first protostar.  Asterisks mark protostars with masses exceeding $10^3~M_\odot$.  The simulation is terminated at $t = 2~\mathrm{Myr}$, corresponding to the typical lifetime of massive stars.  Even at the end of the simulation, the stars remain embedded within dense circumstellar and circumbinary disks.  
(c–e) Time evolution of the stellar mass (panel~c), accretion rate (panel d), and stellar radius (panel e) for the four most massive protostars.  The stellar masses grow steadily with time, reaching final masses of several $10^5~M_\odot$.  Discrete jumps in the mass evolution correspond to stellar mergers, which are accompanied by sharp increases in the accretion rate.  The stellar radii expand to $100$–$1000~R_\odot$ as a result of the high mass accretion rates of $0.01$--$1~M_\odot\,\mathrm{yr^{-1}}$ \citep{Hosokawa+2013}.
}\label{fig3:protostellar_evolution}
\end{figure}

The progenitor cloud does not form a single collapsing object, but instead undergoes hierarchical fragmentation into a dense multiple system whose members experience sustained supergiant phases that suppress radiative feedback.
To examine the feasibility of forming extremely massive seed BHs, we performed a further high-resolution radiation-hydrodynamic simulation for the progenitor cloud of MBH1 that resolves gas densities up to $10^8~\mathrm{cm^{-3}}$.
Figs.~\ref{fig3:protostellar_evolution}(a,b) show the gas density distribution during the protostellar evolution.
Each protostar accretes mass at a rate of $0.1$–$1~M_\odot\,\mathrm{yr^{-1}}$, and the most massive one reaches $\sim4\times10^5~M_\odot$ about $2~\mathrm{Myr}$ after the formation of the first protostar (panels c and d).
Because of these high accretion rates, the stars remain in an inflated supergiant phase, thereby weakening radiative feedback and allowing continued growth \citep{Hosokawa+2013}.
Fig.~\ref{fig3:protostellar_evolution}(e) shows that the stars mostly stay in this supergiant phase, with radii larger than $100\,R_\odot$.
Due to the bursty nature of accretion, their radii also oscillate with time, repeating cycles of Kelvin–Helmholtz contraction and re-inflation driven by entropy injection.
This behavior greatly reduces the UV emissivity and prevents photoevaporation, keeping gas available for subsequent BH growth.
The stars collapse into BHs roughly $2~\mathrm{Myr}$ after their formation, marking the end of the high-resolution simulation.
At the time of collapse, dense gas remains tightly bound around the stars.
These complementary calculations confirm that rapid accretion and limited feedback naturally lead to the formation of supermassive stars that ultimately collapse into heavy BH seeds.

\vspace{0.5em}
\noindent{\textbf{Spectral signatures of circum-BH disks}}

Dense circum-BH disks in our simulation naturally reproduce the physical conditions required to explain the V-shaped continua and Balmer-series absorption features observed in LRDs \citep{Matthee+2024, Hviding+2025, de_Graaff+2025, Rusakov+2025}. Gas at densities $n \gtrsim 10^8~\mathrm{cm^{-3}}$ enhances the $n=2$ population of hydrogen, producing strong Balmer absorption \citep{Ferland+2017, Inayoshi+2025}. Such high densities imply large electron column densities and hence Thomson optical depths sufficient to broaden the H$\alpha$ emission line to widths exceeding $1000~\mathrm{km\,s^{-1}}$, consistent with observations \citep{Rusakov+2025, Chang+2025, Torralba+2025}. Importantly, the H$\alpha$ emission arises directly from the dense gas around the BHs without requiring an externally imposed broad-line region, closely resembling the recently proposed “BH-star” systems \citep{Inayoshi+2025, Naidu+2025, Torralba+2025, de_Graaff+2025, Kido+2025}.

\begin{figure}[h]
\centering
\includegraphics[width=1.0\textwidth]{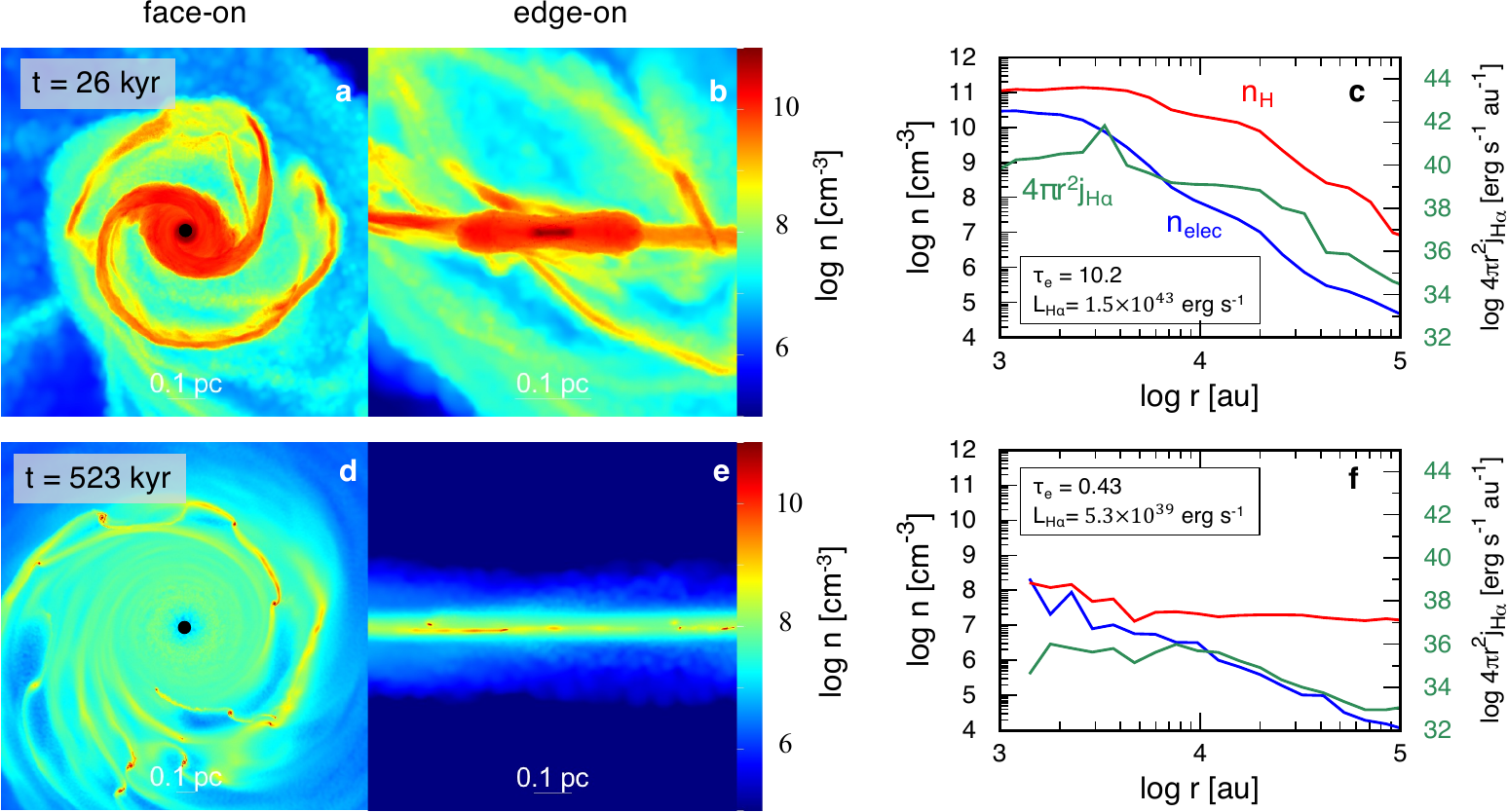}
\caption{ 
Formation and evolution of the dense gas disk around MBH2.
(a,d) Face-on views of the dense circum-BH disk at $t=26$ and $523~\mathrm{kyr}$, showing its rapid emergence and subsequent evolution.
(b,e) Edge-on views at the same epochs, illustrating a vertically thick disk at early times that evolves into a thin, rotationally supported structure.
(c,f) Radial profiles of the hydrogen number density (red), electron number density (blue), and H$\alpha$ emissivity from gas shells at the corresponding radii (green). Only gas within $30^\circ$ of the disk plane is included.
The estimated H$\alpha$ luminosity and the Thomson-scattering optical depth at $r=10^4~\mathrm{au}$ are indicated.
}\label{fig4:BH_star}
\end{figure}

To compare our heavy-seed BH system with the BH-star interpretation, we carry out a dedicated zoom-in simulation for MBH2, restarting at the moment of seed formation and resolving the circum-BH environment down to $500~$au. Fig.~\ref{fig4:BH_star} shows snapshots at 26 and 523 kyr after BH formation. Shortly after formation, the BH is embedded in a dense, geometrically thick disk with hydrogen densities exceeding $n_{\rm H} > 10^{10}\,\mathrm{cm^{-3}}$, sufficient to maintain a large $n=2$ population (Fig.~\ref{fig4:BH_star}c). The resulting H$\alpha$ luminosity within $r < 10^4$~au is $L_{\mathrm{H\alpha}} = 1.5\times10^{43}\,\mathrm{erg\,s^{-1}}$, which is comparable to values inferred for LRDs \citep{Harikane+2023, Matthee+2024}. The Thomson optical depth reaches $\tau_e = 10.2$ at this radius, large enough to generate strong broadening. Although this exceeds observationally inferred values of $\tau_e \sim 0.1$–$1$ \citep{Rusakov+2025}, the effective optical depth for H$\alpha$ photons is smaller because most of the emission originates near the outer disk ($r \sim 10^4$~au). In this early phase, the heavy-seed BH system therefore satisfies the physical conditions associated with the spectral features of a BH-star.

The obscured phase ends after several hundred kyr. At 523 kyr, the BH remains surrounded by dense gas ($n_{\rm H} \gtrsim 10^8\,\mathrm{cm^{-3}}$) with a moderate Thomson depth of $\tau_e = 0.43$, but the H$\alpha$ luminosity declines to $5.3\times10^{39}\,\mathrm{erg\,s^{-1}}$, below typical LRD values (Fig.~\ref{fig4:BH_star}f). Producing strong H$\alpha$ emission in this later stage requires additional line emission from the inner ($\lesssim 100$~au) region, analogous to a compact broad-line region. The accretion rate remains near the Eddington limit, yielding an inner disk luminosity of $\sim10^{44}\,\mathrm{erg\,s^{-1}}$, of which $0.1$–$1\%$ is expected to emerge in H$\alpha$ \citep{Greene+2005}. Thomson scattering in the outer disk then redistributes this radiation and produces broadened emission, yielding H$\alpha$ luminosities of $10^{41}$--$10^{42}\,\mathrm{erg\,s^{-1}}$, consistent with AGNs. These results suggest that the system can naturally transition from an LRD-like, heavily obscured phase into a more AGN-like, less obscured accretion state on timescales of $\sim 0.1$--$1$ Myr following the bursty accretion episode.

\vspace{0.5em}
\noindent{\textbf{Implications for early BH populations}}

Our results demonstrate that heavy-seed formation is a natural outcome of early structure formation in overdense regions, with profound implications for the observed abundance of early quasars and their gravitational-wave signatures.
The overmassive BHs produced in these simulations represent plausible progenitors of the first quasars, bridging the gap between the birth of seed BHs and the luminous active nuclei observed by JWST.

Our model shows that BHs can rapidly grow to $\sim10^{7}~M_\odot$ through the combination of direct collapse and subsequent super-Eddington accretion.
This configuration -- a massive BH embedded in a radiation-pressure-supported, optically thick envelope and fed at highly super-Eddington rates -- closely resembles the ``quasistar'' solutions proposed in analytic models 
\citep[e.g.][]{Begelman+2008, Volonteri2010, Kido+2025, Santarelli+2025}. 
Our simulations provide a fully cosmological realization of such quasistar-like systems, and follow their formation, migration, and subsequent growth within a hierarchically assembling galaxy.
This early phase of rapid growth helps overcome the long-standing difficulty that canonical massive BH seeds ($\sim10^{5}~M_\odot$) cannot efficiently accrete from hot, feedback-heated gas \citep{Dubois+2012, Latif+2018, Chon+2021}.
While our current simulation does not yet resolve the full galactic gas inflows required to sustain long-term Eddington accretion, it suggests that, once such large-scale supply is captured, the BHs could continue growing to $\sim10^{9}~M_\odot$, accounting for the luminous quasars already observed at $z\sim7$ \citep{Mortlock+2011, Banados+2018, Wang+2021}.

Although the super-Eddington accretion phase is short compared with the cosmic age, multiple heavy seeds form nearly simultaneously in the vicinity of massive halos, each experiencing a brief period of super-Eddington growth.
The presence of several such sources increases the likelihood of observing galaxies in an active, LRD-like phase.
Since our calculation covers only a portion of the regions surrounding massive halos, the true occurrence rate of heavy-seed formation may be even higher than inferred here.
As these BHs continue to merge during hierarchical galaxy assembly, they are expected to generate strong gravitational-wave signals detectable by the \textit{Laser Interferometer Space Antenna} (LISA).
Extrapolating to cosmological scales, the predicted abundance of heavy seeds can account for the luminous quasar population observed by JWST at $z>7$.
Their subsequent coalescences contribute gravitational-wave events with characteristic strains of $h_c\sim10^{-17}$–$10^{-16}$ at millihertz frequencies \citep{DeGraf+2024}.
Heavy-seed scenarios similar to ours therefore imply LISA event rates of tens to hundreds over a 3-year mission \citep{Sesana+2007, McCaffrey+2025}.
Taken together, these findings suggest that the early Universe hosted numerous short-lived, rapidly growing BHs that profoundly influenced the assembly of the first massive galaxies.

\clearpage

\phantomsection
\noindent\textbf{Methods}

Our cosmological initial conditions are based on the simulation Phi-4096 of \citep{Ishiyama+2021}. 
In this work, a dark-matter-only $N$-body simulation was performed in a comoving box of side length $16~h^{-1}\mathrm{Mpc}$. 
The initial condition is generated by \textsc{MUSIC} \citep{MUSIC} at a redshift of $127$.
The adopted cosmological parameters follow the latest measurement by \textit{Planck} \citep{Planck2020}: $\Omega_\mathrm{m}=0.31$, $\Omega_\Lambda=0.69$, $\Omega_\mathrm{b}=0.048$, $h=0.68$, $n_\mathrm{s}=0.96$, and $\sigma_8=0.83$. 
The simulation employed $4096^3$ dark matter particles, corresponding to a particle mass of $5.13\times10^3~h^{-1}M_\odot$. 
This high resolution allows the identification of mini-halos with masses down to $10^5~h^{-1}M_\odot$, sufficient to resolve the sites of Population III (Pop III) star formation and to track chemical enrichment across cosmic time.
Halo merger trees were constructed from simulation snapshots between $z=35$ and $z=7.5$ using the \textsc{\small ROCKSTAR} phase space halo/subhalo finder \citep{Behroozi+2013} and the \textsc{\small CONSISTENT-TREES} merger tree code \citep{Behroozi2013b}. 
On top of these, a semi-analytic galaxy formation model was implemented to follow gas cooling, Pop II and Pop III star formation, supernova feedback, metal enrichment, and local Lyman-Werner (LW) radiation fields \citep{Ishiyama+2025}. We use the result of their model with no baryon streaming motion ($0\sigma_\text{vbc}$).

The details of the semi-analytic model largely follow those described in \citet{Agarwal+2012} with recent updates of the treatment of early star formation \citep{Visbal+2020}.
Star formation in Pop II halos follows the standard prescription used in galaxy formation models, in which the cold gas component is converted into stars on a timescale $t_\mathrm{SF}=t_\mathrm{dyn}/\alpha_*$ with efficiency $\alpha_*=0.03$ \citep{Kauffmann+1993}. 
Along with star formation, the model assumes that the rate at which cold gas is reheated into the hot gas phase by associated SN feedback is proportional to the star formation rate, $\gamma \dot{M}_*$, where $\gamma = (V_\text{halo}/110~\mathrm{km~s^{-1}})^{-1.74}$ \citep{Cole+2000, Agarwal+2012}.
When the progenitor halos never formed any stars, the stellar population follows a Pop III initial stellar mass function (IMF). 
The characteristic Pop III stellar mass is determined by the halo mass growth rate, which correlates with the final stellar mass \citep{Hirano+2015}.
Stellar populations emit LW radiation that photodissociates molecular hydrogen and delays star formation in nearby halos. 
The LW radiation intensity is calculated separately for Pop~II and Pop~III stars \citep{Johnson+2013}:
\begin{align}
J_{21,\mathrm{III}} &= \sum_i 15 \left( \frac{r_i}{1~\mathrm{kpc}} \right)^{-2}
                     \left( \frac{M_{\mathrm{PopIII},i}}{1000~M_\odot} \right), \\
J_{21,\mathrm{II}}  &= \sum_i 3 \left( \frac{r_i}{1~\mathrm{kpc}} \right)^{-2}
                     \left( \frac{M_{\mathrm{PopII},i}}{1000~M_\odot} \right),
\end{align}
where $r_i$ is the distance to halo $i$, and $M_{\mathrm{PopIII},i}$ and $M_{\mathrm{PopII},i}$ are the masses of Pop~III and Pop~II stars formed within the past $5~\mathrm{Myr}$ in halo $i$, respectively. 
The sums run over all halos that host Pop~III or Pop~II star formation. 
The coefficient in the Pop~II expression is derived assuming a Scalo IMF \citep{Scalo1998} and a stellar metallicity of $Z=0.001$ \citep{Barkana+2001, Greif+2006}. 

The LW background is primarily contributed by Pop~II stars, and its amplitude depends on assumptions regarding the IMF and the stellar models adopted. 
We note that uncertainties in the LW-intensity modeling have only a minor impact on our results, as we later test explicitly. 
A halo begins to form Pop~III stars once its mass exceeds the critical threshold determined by the local LW intensity \citep{Kulkarni+2021}.
Under strong LW irradiation, star formation is delayed until the halo virial temperature reaches $\sim8000$ K, at which point Ly$\alpha$ cooling triggers rapid collapse. 
Such halos typically experience high mass accretion rates and host more massive Pop III stars, extending to supermassive stars (SMSs) with $M_\ast\sim10^5~M_\odot$, which subsequently collapse into heavy black-hole (BH) seeds. 
Indeed, \citet{Ishiyama+2025} identified more than $\sim10^4$ SMSs with the masses larger than $10^5~M_\odot$ as heavy seed candidates in their simulation volume.

From this dataset, we select one representative halo predicted to form a massive seed, which serves as the target of our follow-up radiation-hydrodynamic simulations described in this work. 
The halo is chosen according to two criteria: 
(1) the local LW intensity at the time of seed formation exceeds the critical value of $J_{21,\mathrm{crit}} = 1000$, 
and (2) the halo is not tidally disrupted by nearby massive galaxies.
Here, $J_{21}$ denotes the far-ultraviolet (FUV) intensity normalized to $10^{-21}~\mathrm{erg\,s^{-1}\,Hz^{-1}\,cm^{-2}\,sr^{-1}}$.
The second condition is necessary because candidate halos are often located near luminous galaxies that provide strong LW irradiation, but their gravitational collapse may be suppressed by external tidal fields \citep{Chon+2016}. 
To exclude such cases, we select halos whose distances from the nearest massive galaxy satisfy $r_\mathrm{dist} > 10\,d_\mathrm{tidal}$, 
where $d_\mathrm{tidal}$ is defined as the separation at which the tidal radius imposed by the neighboring galaxy equals the virial radius of the candidate halo. 
Within the simulated volume, we identify 60 halos that meet both criteria. 
Among them, we focus on the one that exhibits the most rapid mass growth, reaching a virial temperature of $T_\mathrm{vir}\simeq 8\times10^{3}\,\mathrm{K}$, which marks the onset of atomic cooling and subsequent collapse.

We trace back the initial position of the selected candidate halo to the cosmological initial conditions at redshift $z=127$. 
Extended Data Fig.~\ref{figex1:Mhalo_evo} shows the time evolution of the halo mass, in which MBH1 forms. The white asterisk marks the redshift at which the virial temperature of the halo reaches $8000~\mathrm{K}$, when it is identified as a DCBH halo.
A zoom-in region is defined to be $40$ times larger than the Lagrangian radius of the target halo, corresponding to a comoving size of $\sim400$~kpc. 
We confirm that this region is sufficiently large to encompass all material relevant to the formation of the heavy seed BH in the selected halo. We measured the overdensity of the specified zoom-in region by computing the mean density within its bounding box of $\sim 400~$ comoving kpc and found it to correspond to a $3.8\sigma$ fluctuation.

\vspace{1.em}
\noindent\textbf{Radiation hydrodynamic calculation}

The radiation-hydrodynamic calculation is performed within this zoom-in region using the moving-mesh code \textsc{\small AREPO} \citep{Springel2010}, extended to include prescriptions for star formation and BH accretion physics. 
The resolution of the zoom-in region at the initial snapshot is identical to that used in \citet{Ishiyama+2021}, where the dark-matter particle and baryonic cell masses are $4.33\times10^3$ and $7.94\times10^2~h^{-1}M_\odot$, respectively. 
At this resolution, star formation within mini-halos can be reliably resolved. 
Adaptive mesh refinement is applied whenever the local cell size falls below $16$ times the local Jeans length, ensuring that gravitational collapse is properly captured and that artificial fragmentation is avoided \citep{Truelove1997}. 
Unlike the semi-analytic model of \citet{Ishiyama+2025}, the formation and properties of primordial stars are directly followed by the high-resolution radiation-hydrodynamic simulations described here. 
Uncertainties in the stellar mass and the resulting FUV background intensity in the semi-analytic model may affect whether a heavy-seed BH forms or not, but we later show that these uncertainties do not significantly alter our main conclusion regarding the rapid emergence of overmassive BHs.

We solve a non-equilibrium primordial chemical network consisting of eight species: $\mathrm{e^-}$, H, $\mathrm{H^+}$, $\mathrm{H_2}$, $\mathrm{H^-}$, D, $\mathrm{D^+}$, and HD. 
The network and associated cooling processes follow the implementation of \citet{Matsukoba+2022}, and include H$_2$ ro-vibrational line cooling, atomic hydrogen line cooling (including Ly$\alpha$), and free-free and free-bound emission of H and $\mathrm{H^-}$. 
The ionization of H, the dissociation of H$_2$ by LW radiation via the Solomon process, and the photodetachment of H$^-$ are also taken into account. The chemical network and adopted reaction rates are similar to those in the minimal chemistry model presented by \citet{Glover+2015}, except that we neglect H$_2^+$ and He.
H$_2^+$ can catalyse H$_2$ formation in primordial gas, but it is less important in environments exposed to the background radiation field assumed in this study, namely a $10^5~\mathrm{K}$ blackbody spectrum \citep{Glover+2015}.
Heating and chemical reactions induced by FUV, extreme-UV (EUV), and X-ray radiation from nearby stars and BHs are also included, as described later. 

Helium chemistry becomes important at gas temperatures of several $10^4$--$10^5~{\rm K}$, where it provides additional cooling channels and increases the free-electron fraction. Although we do not solve the full non-equilibrium helium chemistry, we estimate the ionization state of helium by assuming chemical equilibrium and include the line-cooling rates of He I and He II. This effect is relevant for gas irradiated by both Pop III stars and accreting BHs.
Radiation from the accreting BH raises the gas temperature well above $10^4~{\rm K}$. The elevated temperature already increases the electron fraction and thereby promotes H$_2$ formation; including non-equilibrium helium chemistry would further enhance the H$_2$ abundance to some extent. However, this does not affect our main conclusion. Even if the enhanced H$_2$ abundance lowers the cloud temperature, the large-scale inflow can be sustained. This is demonstrated by the formation of MBH2, where a heavy seed BH forms under a massive inflow despite efficient H$_2$ cooling that lowers the gas temperature below $1000~{\rm K}$ (see Extended Data Figure~\ref{figex6:2dhist}). 

To avoid numerical contamination from the coarse outer region, gas cells outside the zoom-in volume are kept at fixed resolution, and both radiative cooling and refinement are disabled.
In this calculation, we treat the FUV radiation from the source halos separately from the radiation emitted by stars and BHs formed in the target region. Here, we define source halos as halos that have already been labelled as star-forming when the target halo satisfies the DCBH formation criteria in the semi-analytic model. These halos host star-forming galaxies and provide the strong FUV radiation required for DCBH formation.
Instead of explicitly following the star formation within the galaxies, we adopt a time-dependent LW intensity derived from the semi-analytic model of \citet{Ishiyama+2025}, which self-consistently evolves the star formation and feedback in the same cosmological volume. The resulting LW flux is applied uniformly to all gas cells within the zoom-in region as a uniform background field. 
We separately solve the ionizing radiation and X-rays emitted by stars and BHs formed inside the zoom-in region, but outside the source halos, using the ray-tracing scheme implemented in \textsc{\small RSPH} \citep{Susa2006, Chon+2017}. We include photoionization of H and the associated photoheating, photodissociation of H$_2$ and HD, and photodetachment of H$^-$. The photodetachment rate of H$^-$ is calculated under the optically thin approximation, whereas the other photoreaction rates are calculated using \textsc{\small RSPH}. We do not include the secondary ionization and heating. The luminosities and spectra of these sources are described later in Section~``Modeling Pop III stars and BHs''.

Extended Data Fig.~\ref{figex2:J21_evo} shows the time evolution of the LW intensity contributed by nearby source galaxies. The intensity rises rapidly and exceeds $J_{21}=1000$ around redshift $z\simeq18$, the critical level required to suppress H$_2$ cooling and enable heavy-seed BH formation \citep[e.g.,][]{Shang+2010, Sugimura+2014, Latif+2014, Regan+2016}. 
The spectrum of the background radiation is modeled as a blackbody with an effective temperature of $T_\mathrm{BB}=10^5$~K, which reproduces the relative rates of H$_2$ photodissociation and H$^-$ photodetachment expected from young star-forming galaxies \citep{Sugimura+2014}. 
We consider the self-shielding of the background LW radiation using the prescription described in \citet{Wolcott-Green+2011}.
We assume that the cloud extends over the local Jeans length, $l_\mathrm{Jeans}$, and estimate the H$_2$ column density as $N_{\mathrm{H}_2} = n_{\mathrm{H}_2} l_\mathrm{Jeans}$, where $n_{\mathrm{H}_2}$ is the number density of H$_2$ in the cell.
We neglect the contribution of ionizing photons from neighboring galaxies, because their mean free path is short and they are strongly attenuated by the surrounding intergalactic medium \citep{Chon+2017}.

During the radiation-hydrodynamic simulation, we do not assume any specific IMF, but instead resolve individual star formation events directly using a sink-particle method. 
A sink particle is introduced once the local gas density exceeds $2\times10^6~\mathrm{cm^{-3}}$, with an accretion radius set to ten times the local mesh size. 
The sink accretes surrounding gas following the local gravitational potential, and its properties are updated accordingly at each time step.
We allow mergers of the sink particles once the distance of two sink particles becomes smaller than the sum of the sink radii.

\vspace{1.2em}
\noindent\textbf{Modeling Pop III stars and BHs}

The luminosity and effective temperature of each protostar are determined from its instantaneous mass and accretion rate by interpolating results from one-dimensional stellar evolution models assuming constant accretion rates \citep{Hosokawa+2009}. 
If the stellar accretion rate exceeds the critical accretion rate of $\dot{M}_\text{crit} \equiv 0.02 ~M_\odot~\mathrm{yr}^{-1}$, we assume the stellar radius expands due to the injection of the large amount of entropy into the stellar envelope \citep{Hosokawa+2013, Haemmerle+2018, Nandal+2023, Nandal+2025}.
During this phase, we assume the surface temperature of the star to be $6000~$K and the luminosity to be the Eddington value following the results of detailed stellar evolution calculations \citep{Hosokawa+2012}.
After the accretion rate becomes smaller than $\dot{M}_\text{crit}$, we gradually shrink the stellar radius to the radius expected for main-sequence stars on the timescale of the surface Kelvin-Helmholtz (KH) time, which is given by ten times the stellar KH time \citep{Sakurai+2015}.
Once the stellar age exceeds the lifetime predicted by these models, the star is converted into a BH particle if its final mass exceeds $260~M_\odot$ \citep{Heger+2002}. 
We note here that all the stars formed during the simulation (the progenitor stars of LBH, MBH1, and MBH2) have masses above $260~M_\odot$ and collapse into the BHs without any supernova feedback. 
The mass accretion rate is measured by the sink method, where the gas inside the sink radius is assimilated and added to the mass of the star particle.
The sink radius is set to be 100 times the cell size, which will be converted into the sink particle, which ranges from $1$--$20~$pc.
Because the sink radius is much larger than the physical stellar radius, we estimate the final stellar mass from the sink accretion rate using the empirical relation derived from previous high-resolution simulations of Pop~III star formation \citep{Hirano+2015, Toyouchi+2023},
\begin{align}
M_* &= 250~M_\odot \left ( \frac{\dot{M}_*}{2.8\times10^{-3}~M_\odot\mathrm{yr^{-1}}} \right )^{0.7},
\end{align}
where $\dot{M}_*$ denotes the time-averaged accretion rate onto the growing protostar. 
When the stellar age reaches its lifetime, any excess gas mass accreted by the sink that exceeds the estimated stellar mass is returned to the surrounding gas cells to conserve mass. 
Throughout the simulation, we assign the sink-particle mass as the instantaneous stellar mass when evaluating radiative feedback. 
This approach somewhat overestimates the stellar luminosity and gives stronger feedback effects compared to the actual stellar mass, so the final stellar mass represents a lower limit to the actual value.

We assume a stellar lifetime of $2~\mathrm{Myr}$, typical for very massive stars with masses above $100~M_\odot$ \citep{Hosokawa+2013}. 
After the stellar age exceeds $2~\mathrm{Myr}$, we convert the stars into BHs. 
However, the lifetime of such stars -- particularly supermassive stars -- remains uncertain. 
Stars with masses above a few $\times 10^5~M_\odot$ become unstable due to general-relativistic effects and collapse into BHs during the hydrogen-burning phase \citep{Fuller+1986, Shibata+2002, Hosokawa+2013, Woods+2017, Uchida+2017, Nandal+2024}. 
This implies that collapse may occur earlier than the canonical lifetime assumed for massive stars. 
Recent calculations \citep{Umeda+2016} show that the threshold mass for general relativistic instability depends on the mass accretion rate, spanning $2$–$8\times10^5~M_\odot$. 
If the stars collapse into BHs earlier, the resulting BHs would remain embedded in dense gas for a longer period, thereby extending the phase of super-Eddington growth.

After the stars collapse into BHs, we convert the corresponding sink particles into BH particles (LBH, MBH1, and MBH2). 
We keep their original accretion radii if the accretion radius is larger than the Bondi radius of the BHs for the ionized gas, otherwise the accretion radii are set to be the Bondi radii. 
Since the typical accretion radius is around a parsec, the simulations resolve the Bondi radius for most of the massive seed BHs, allowing us to  directly follow the gravitational capture of gas onto the BHs. 
The spectral energy distribution (SED) of accreting BHs is modeled as a double power-law continuum, 
$F_\nu \propto \nu^{-0.6}$, extending from $1~\mathrm{eV}$ to $10~\mathrm{eV}$, and $F_\nu \propto \nu^{-1.5}$, extending from $10~\mathrm{eV}$ to $1~\mathrm{keV}$ \citep{Sazonov+2004}. 
This represents the integrated emission from the accretion disk and its associated corona. 
We allow the accretion rate to exceed the classical Eddington limit, consistent with theoretical models of super-Eddington accretion flows \citep[e.g.,][]{Begelman1978, McKinney+2014, Inayoshi+2016, Sadowski+2016}. 
We assume the slim disk solution to give a bolometric luminosity when the accretion rate is higher than the Eddington rate,
\begin{align}
L_\mathrm{bol} = \left \{ \begin{array}{ll} \epsilon_\mathrm{r}\, \dot{M}_\mathrm{acc}\, c^2 & (\dot{M}_\text{acc} \le 2\dot{M}_\text{Edd} ),  \\
2 \left [ 1 + \log \left (\frac{\dot{M}_\mathrm{acc}}{2\dot{M}_\text{Edd}} \right ) \right ] L_\text{Edd} & (\dot{M}_\text{acc} > 2\dot{M}_\text{Edd} ),
\end{array} \right.
\end{align}
where $\epsilon_\mathrm{r}=0.1$ is the radiative efficiency, $\dot{M}_\mathrm{acc}$ is the instantaneous gas accretion rate onto the BH, $\dot{M}_\text{Edd}$ is the Eddington accretion rate, and $L_\text{Edd} \equiv \dot{M}_\text{Edd} / \epsilon_\text{r} c^2$ is the Eddington luminosity \citep{Watarai+2001}.
The emitted radiation is coupled to the surrounding gas through photoionization heating, which regulates the accretion flow and drives intermittent outflows around the BHs.
We also allow mergers of BHs once the distance between two BHs becomes smaller than 10 physical parsec.

We do not include kinetic feedback from accreting BHs, such as jets or winds. Recent GRMHD simulations have shown that BHs accreting at super-Eddington rates can launch magnetically arrested jets and disk winds. If such jets break out and transport mass and momentum to larger scales in the surrounding medium, the accretion efficiency may decrease below the Eddington value \citep{Regan+2019, Massonneau+2023}.
The efficiency of jet launching should depend on the BH spin and magnetic field strength, both of which are highly uncertain. This uncertainty arises because modelling them would require following the stellar rotation at the birth of DCBHs and the detailed magnetic dynamo processes during accretion. The jet opening angle and its interaction with the accreting gas are also uncertain.
We therefore regard our simulation as a fiducial model that provides an upper limit on BH growth under the assumption that kinetic feedback is absent.

\vspace{1.em}
\noindent\textbf{Zoom-in calculation of seed BH formation in the isolated cloud}

To assess the feasibility of heavy-seed BH formation, we perform a higher-resolution follow-up simulation that resolves the individual protostars expected to be the progenitors of the seed BHs. 
We focus on the formation site of the first heavy seed BH, MBH1, identified in the cosmological radiation-hydrodynamic run. 
This seed, with a final mass of $6\times10^{5}\,M_\odot$, forms at redshift $z\simeq14$. 
We extract a cubic region of $1~\mathrm{comoving~kpc}$ on a side centered on the host halo, shortly before the onset of collapse when the central gas density reaches $10^{3}~\mathrm{cm^{-3}}$. 
At this point, the central region of the cloud, of size a few $10~$pc, becomes Jeans-unstable (see Extended Data Fig.~\ref{figex5:Menc_MBH}). 
The re-simulation is performed using the same \textsc{\small AREPO} framework, but with enhanced spatial and mass resolution to capture the internal fragmentation of the collapsing gas cloud. 
Sink particles are introduced once the local gas density exceeds $2\times10^{8}~\mathrm{cm^{-3}}$, representing the formation of individual protostars. The sink radius is set to be four times the cell size, which will be converted into the sink particle, which ranges from $3000$--$8000~$au.
We allow mergers of the sink particles once the distance of two sink particles becomes smaller than the sum of the sink radii.

The numerical resolution adopted here is insufficient to resolve the opacity limit at $n \sim 10^{16}~\mathrm{cm^{-3}}$, above which the gas becomes optically thick and protostars form \citep{Omukai2001}. Indeed, the physical sizes of growing protostars are expected to be $\sim 0.05$--$100~\mathrm{au}$, depending on the stellar mass and accretion rate \citep{Hosokawa+2012}, which is below our numerical resolution.
We can nevertheless compare our results with higher-resolution studies. \citet{Becerra+2018} studied the protostellar evolution changing $n_\mathrm{adib}$, the density above which the gas becomes adiabatic, from $10^8~\mathrm{cm^{-3}}$ to $10^{12}~\mathrm{cm^{-3}}$. 
They have found that the mass evolution of supermassive protostar is not significantly affected by the numerical resolution, because the growth rate is mainly regulated by the large-scale accretion flow. They followed the early stellar evolution for $10^4$--$10^5~\mathrm{yr}$ and found that the accretion rate remains at $\sim 1~M_\odot~\mathrm{yr^{-1}}$, comparable to our results.
Similarly, \citet{Chon+2025} followed protostellar evolution in a DCBH halo for $2~\mathrm{Myr}$ with $n_\mathrm{adib}=10^{11}~\mathrm{cm^{-3}}$. They found that the initial mass accretion rate remains around $1~M_\odot~\mathrm{yr^{-1}}$. 
In the simulation of \citet{Chon+2025}, however, the accretion rate drops below $0.01~M_\odot~\mathrm{yr^{-1}}$ at $10^4$--$10^5~\mathrm{yr}$ after protostar formation because of the limited gas supply. This contrasts with our results, where efficient accretion continues until $\sim 2~\mathrm{Myr}$ after protostar formation.

\vspace{1.em}
\noindent\textbf{Zoom-in calculation of the BH accretion phase}

To reproduce the density structure around the massive seed BH, we performed a high-resolution simulation that resolves gas densities up to $10^{12}~\mathrm{cm^{-3}}$ and spatial scales down to $500~\mathrm{au}$ around the central BH.
We take the snapshot at the moment when MBH2 forms at $z=13.1$ and follow the evolution for $30~\mathrm{kyr}$ after its emergence.
To capture the later evolution up to $0.5~\mathrm{Myr}$ after MBH2 formation, we also run a complementary simulation with a larger sink radius of $5000~\mathrm{au}$.
Starting from the snapshot at $0.48~\mathrm{Myr}$, we then reduce the sink radius back to $500~\mathrm{au}$ and continue the calculation to follow the detailed structure of the circum-BH disk for an additional $0.55~\mathrm{kyr}$.
This procedure allows us to track both the long-term disk evolution and the small-scale accretion flow onto the BH.

\vspace{1.em}
\noindent\textbf{Host halo evolution of MBHs} \\
Extended Data Fig.~\ref{figex7:migration} shows the redshift evolution of the distance between the source galaxy and the host halos of MBH1, MBH2, and LBH. All BHs in our simulation form at distances of about $10~\mathrm{physical\ kpc}$ from the source galaxy, outside the virial radius of the source halo (dashed line). After their formation, the BHs migrate toward the source galaxy and enter the virial radius of the source halo at $z\sim 12$.
Initially, the BHs follow eccentric orbits, as indicated by the oscillatory behavior of their distances. Over time, however, their orbits gradually decay, and the BHs settle toward the galaxy center by around $z \sim 8$. We also find that, around this epoch, their accretion rates rise to $\sim 0.1$--$1$ times the Eddington value (Fig.~\ref{fig1:snapshots}f).

The host halos of the BHs may avoid metal enrichment because they are well separated, by $\sim 10$ physical kpc, from the source galaxy that drives winds and pollutes the surrounding IGM with metals. Analytic estimates of galactic-wind expansion by \citet{Dijkstra+2014} show that such winds expand only to $\sim 1$--$2$ physical kpc within $300~\mathrm{Myr}$. Similarly, \citet{Ventura+2024} statistically studied the distribution of superbubbles around star-forming halos in a $10~h^{-1}\mathrm{Mpc}$ cosmological region and found that the bubble size is at most $\sim 5~\mathrm{kpc}$. These scales are much smaller than the separation found in our simulation, suggesting that the BH host halos can remain metal-poor despite their proximity to the source galaxy.

We have shown that the key to forming an overmassive BH population is the formation of DCBHs in massive halos, with virial temperatures reaching $\sim 40000~\mathrm{K}$. We attribute this to the delayed onset of protostar formation after Ly$\alpha$ cooling becomes effective, caused by the finite collapse timescale of the halo and by dynamical heating.
Extended Data Fig.~\ref{figex8:host_halo_evolution} shows the time evolution of the halo growth rate and the peak gas density in the host halo of MBH1. The gas density increases steadily, but protostar formation (at $300~$Myr) occurs only $\sim 90~\mathrm{Myr}$ after the halo virial temperature first reaches $8000~\mathrm{K}$ (at $210~$Myr), which is often assumed to mark the onset of DCBH formation.
This delay arises for two reasons. First, the collapse timescale is of the order of the free-fall time,
\begin{align}
t_\mathrm{ff} \sim \sqrt{\frac{1}{G\rho}}
= 27.4~\mathrm{Myr}
\left( \frac{n}{10~\mathrm{cm^{-3}}} \right)^{-1/2},
\end{align}
where $n$ is the gas density of the gravitationally unstable region. This implies that DCBH formation should occur only after a timescale of order $\sim 10~\mathrm{Myr}$ once the halo virial temperature exceeds $8000~\mathrm{K}$.
The halo mass can grow during the initial free-fall phase to exceed $T_\text{vir} = 8000~$K.
Second, dynamical heating caused by halo mergers further delays the collapse. During the first $\sim 30~\mathrm{Myr}$ after the virial temperature reaches $8000~\mathrm{K}$, the peak gas density increases only weakly. Around this epoch, the host halo merges with nearby halos or clumps, which deposits additional gravitational and kinetic energy into the collapsing gas and suppresses runaway collapse \citep{Yoshida+2003, Wise+2019}. After these clump mergers subside and the halo growth rate decreases, the peak gas density begins to rise rapidly, eventually leading to the formation of the protostar that later collapses into MBH1. During this delay, the host halo grows sufficiently massive to enable subsequent super-Eddington accretion.

\vspace{1.em}
\noindent\textbf{Formation of extremely massive seed BHs}

Here we show how extremely massive seed BHs form -- through the formation of heavy seeds followed by a brief phase of super-Eddington accretion.
Extended Data Fig.~\ref{figex1:Mhalo_evo} shows the time evolution of the mass of the halo that later hosts the heavy seed BH (MBH1).
At the time of protostar formation, the halo mass is about $2\times10^8~M_\odot$, corresponding to a virial temperature of $4\times10^4$~K.
This is much higher than the canonical virial temperature of $8000$~K at which halo collapse is typically expected.
Indeed, the semi-analytic model predicts that this halo would have collapsed at $z\sim18.1$, when its mass was $\sim10^7~M_\odot$, nearly an order of magnitude smaller than the host halo of MBH1.
During the delayed collapse, more gas accumulates within the halo, leading to the formation of more massive stars and enhanced accretion onto the resulting BHs.

Once the gas within a halo becomes gravitationally unstable, it collapses to form stars. The final stellar and BH masses are determined by how much gas is accreted onto the central protostar during its growth phase. Extended Data Fig.~\ref{figex4:radial_profile} shows the radial profiles of (a) gas density, (b) temperature, and (c) escape velocity at the time of protostar formation. The grey line represents the light-seed case (LBH), while the purple and green lines correspond to the two massive seeds, MBH1 and MBH2, respectively. The density structure differs markedly between light and heavy seeds. In the massive-seed cases, the high-density core extends to radii an order of magnitude larger than in the light-seed case, a consequence of the higher gas temperature in the collapsing cloud. Panel (b) shows that the temperature in heavy-seed formation remains near $10^4$~K -- an order of magnitude higher than in the light-seed case -- due to strong UV irradiation that suppresses molecular cooling. The collapse of this larger, hotter core also raises the escape velocity, which exceeds $10~\mathrm{km~s^{-1}}$. This high binding energy facilitates continued mass accretion, as the ionized gas remains gravitationally bound and can feed the central protostar efficiently.

To visualize how gravitational collapse proceeds, we compare the enclosed mass with the critical Bonnor–Ebert (BE) mass, the threshold above which a cloud becomes gravitationally unstable to collapse \citep{Bonnor1956, Ebert1955}.
The top panels in Extended Data Fig.~\ref{figex5:Menc_MBH} show the BE mass (green) and the enclosed mass (purple) for LBH, MBH1, and MBH2, from left to right.
In the light-seed case, only the innermost $0.1$–$1$~pc region is gravitationally unstable, whereas in the heavy-seed cases the unstable region extends to $10$–$100$~pc.
The bottom panels display the ratio of enclosed to BE mass as a function of enclosed mass; regions where this ratio exceeds unity are gravitationally unstable.
Only gas within $\sim10^3~M_\odot$ becomes unstable in the light-seed case, while up to $\sim10^6~M_\odot$ it is unstable in the heavy-seed cases.
These characteristic mass scales roughly correspond to the resulting BH masses: $\sim800~M_\odot$ for LBH, and $3\times10^5$ and $6\times10^5~M_\odot$ for MBH1 and MBH2, respectively.
The dynamical time at the edge of the gravitationally unstable region is
\begin{align}
t_\mathrm{dyn} \sim \sqrt{\frac{R^3}{G M_\mathrm{enc}}} = 1.5\times10^7~\mathrm{yr} \left(\frac{R}{100~\mathrm{pc}}\right)^{3/2} \left(\frac{M_\mathrm{enc}}{10^6~M_\odot}\right)^{-1/2},
\end{align}
which is longer than the lifetime of massive stars ($\sim2$~Myr).
This implies that the dense envelope remains bound and should continue to fall onto the BHs after the central protostars collapse, unless feedback from the protostars clears it away.

The temperature decrease at $r\lesssim1$~pc in MBH2 is caused by the enhanced electron fraction and the subsequent increase in molecular hydrogen formation, similar to the Pop III star formation in a fossil H~II region \citep{Yoshida+2007}.
Extended Data Fig.~\ref{figex6:2dhist} shows two-dimensional histograms of temperature (left), molecular hydrogen fraction (middle), and electron fraction (right) as functions of gas density, for MBH1 (top) and MBH2 (bottom).
Gas at temperature $T\sim100$–$1000$~K in MBH2 shows an enhanced H$_2$ fraction, indicating that molecular formation and the associated cooling are responsible for the temperature drop.
The electron fraction in the outer envelope is almost fully ionized but decreases toward higher densities as recombination proceeds.
The elevated electron fraction during collapse promotes H$_2$ formation through the H$^-$ channel, the dominant pathway in primordial gas, which is catalyzed by free electrons.
The ionization source is the strong radiation from MBH1, which fully ionizes and heats the surrounding gas.
Indeed, the temperature rise at $r\gtrsim10^3$~pc in MBH2 indicates that intense radiation from MBH1 photoionizes the intergalactic medium (IGM).
Despite the enhanced H$_2$ formation, a massive seed BH still forms in MBH2, as large-scale gravitational instability drives a strong gas inflow that overwhelms the cooling. This behavior is analogous to the heavy-seed formation in the super-competitive accretion scenario, in which fine-structure lines and dust cooling reduce the gas temperature, while large-scale collapse collects a substantial amount of gas to form massive seed BHs \citep{Chon+2020, Chon+2025}.

\vspace{1.em}
\noindent\textbf{Observability of growing MBHs} \\
To estimate the optical depth and H$\alpha$ luminosity, we restrict the analysis to gas cells located within $30^\circ$ of the disk plane.
We estimate the electron column density, $N_\mathrm{e}$, by summing the total number of electrons contained within each spherical shell and dividing by the shell’s surface area, yielding an average column density at radius $r$.
The Thomson optical depth is then calculated as $\tau = N_\mathrm{e}\,\sigma_\mathrm{T}$, where $\sigma_\mathrm{T} = 6.65\times10^{-25}~\mathrm{cm^{2}}$ is the Thomson scattering cross section.
In estimating the H$\alpha$ luminosity we have assumed case-B recombination. This may underestimate H$\alpha$ luminosity because resonance scattering will increase the H$\alpha$ emissivity at densities of $n \sim 10^{10\text{--}11}\,\mathrm{cm^{-3}}$ \citep{Inayoshi+2022}.
The detailed radiative-transfer calculations for similar conditions indicate that the emergent  H$\alpha$/H$\beta$ is enhanced to values of $\sim 6$–$10$ consistent with the observed values for LRDs \citep{Chang+2025}.

Current Chandra constraints indicate that LRDs are generally X-ray weak. Only three LRDs have secure individual X-ray detections so far \citep{Kocevski+2025, Hviding+2026}, while the remaining LRD population is largely undetected on an individual basis. Stacking analyses have also yielded mostly tentative signals or non-detections, including stacks of X-ray-undetected sources \citep{Ananna+2024, Yue+2024, Sacchi+2025}.
To assess the detectability of X-rays emitted by the central BH in our simulation, we evaluated the hydrogen column density, $N_\mathrm{H}$, at two different snapshots, $t=26$ and $523~\mathrm{kyr}$. Extended Data Fig.~\ref{figex3:NH_distribution} shows the distribution of $N_\mathrm{H}$ measured along 64 different lines of sight evenly spaced around the accreting MBH2.
During the early super-Eddington accretion phase at $t=26~\mathrm{kyr}$, the column density exceeds $10^{26}~\mathrm{cm^{-2}}$ in all directions. This implies that X-rays emitted from the BH are strongly Compton scattered and therefore undetectable from any viewing angle, consistent with the non-detections obtained from stacking Chandra data \citep{Sacchi+2025}.
At the later epoch, nearly Eddington phase at $t=523~\mathrm{kyr}$, the BH remains embedded in dense gas, but the column density decreases to $N_\mathrm{H}=10^{25}$--$10^{26}~\mathrm{cm^{-2}}$, lower than in the early super-Eddington phase. Even in this phase, however, the X-ray emission is still expected to be difficult to observe.
This analysis indicates that the LRD-like system found in our study remains X-ray dark for at least the first $\sim 0.5~\mathrm{Myr}$, in agreement with current observational constraints.

\vspace{1.em}
\noindent\textbf{Inefficient growth of Pop III remnant BH} \\
We find that LBH undergoes almost negligible growth during the simulation. Its mass remains close to $800~M_\odot$ until $z \sim 12$, followed only by modest growth thereafter. We mainly attribute this inefficient growth to the shallow gravitational potential of the host halo at the time of Pop III star formation. Extended Data Fig.~\ref{figex4:radial_profile}c shows that the escape velocity is only $\sim 2$--$5~\mathrm{km~s^{-1}}$, smaller than the sound speed of ionized gas, $\sim 10~\mathrm{km~s^{-1}}$. This indicates that the halo potential is too shallow to retain ionized gas, so the gas inside the halo is rapidly expelled by photoevaporation \citep{Alvarez+2009, Jeon+2012, Latif+2018, Jeon+2023}.
Indeed, the gas density around LBH remains $\lesssim 1~\mathrm{cm^{-3}}$, with a temperature of $\sim 10^4~\mathrm{K}$ (Figs.~\ref{figex4:radial_profile}a and b). Under these conditions, the Bondi accretion rate is only $\sim 10^{-8}$--$10^{-7}~M_\odot~\mathrm{yr^{-1}}$, much lower than the Eddington accretion rate of LBH.

Around $z \sim 12$, the halo potential becomes sufficiently deep for gas to begin accumulating at the halo centre. However, the inflowing gas is predominantly supplied to MBH1 rather than to LBH. MBH1 formed in a halo located close to the original host halo of LBH, and the two host halos subsequently merged. After the merger, MBH1 becomes the primary accretor because of its larger mass, while LBH remains less efficiently fed. 
This is partly because LBH has a smaller mass and therefore a longer dynamical-friction timescale than the massive BHs, further suppressing its orbital decay and growth \citep{Latif+2018, Chon+2021}. 
Although stochastic interactions with nearby gas increase the mass of LBH to several thousand solar masses, its accretion rate remains much lower than those of the massive BHs.

\vspace{1.em}
\noindent\textbf{DCBH Formation in lower-z universe} \\
To test how ubiquitous DCBH formation and LRD-like systems are, and to examine whether this phenomenon can occur in the later Universe, we performed RHD calculations for another candidate region selected from \citet{Ishiyama+2025}. To identify a lower-redshift sample, we chose a candidate halo that satisfies the DCBH formation criteria at $z \sim 14$, the latest formation redshift among our 62 samples.

We find eight seed BHs forming in the region around the target halo. Extended Data Fig.~\ref{figex9:MBH_evolution_ID22}a shows the spatial distribution of the massive BHs formed in this region at $z=7.011$. At this snapshot, the lowest-redshift DCBH, labelled MBH8, has already formed. Some of the BHs have migrated to the centres of massive halos, while other clumps around the target halo are still forming BHs.
Extended Data Fig.~\ref{figex9:MBH_evolution_ID22}b and c show the redshift evolution of the BH mass and the accretion rate normalized by the Eddington rate, respectively. All seed BHs grow beyond $10^6~M_\odot$ after experiencing brief episodes of super-Eddington accretion. Some BHs, such as MBH1 and MBH3, repeatedly increase their accretion rates and undergo multiple super-Eddington phases.
These results demonstrate that DCBH formation followed by later super-Eddington accretion is not confined to the earliest cosmic epochs, but can also occur at later times, overlapping with the redshift range in which LRDs have been detected.

\vspace{1.em}
\noindent\textbf{Robustness tests} \\
To assess numerical robustness and model sensitivity, we performed one higher-resolution run (lv13) and three simulations with reduced FUV background intensities (w2, w5, and w10). 
In the latter three runs, we kept the parameters of the semi-analytic model fixed, but reduced the LW intensity from its fiducial value obtained from the model.
In the higher-resolution run, we adopt an effective resolution of $8192^3$ in the zoom-in region, corresponding to dark-matter and baryonic particle masses of $542$ and $99.2~h^{-1}M_\odot$, respectively. 
This resolution is sufficient to resolve mini-halos in which the first stars form \citep{Yoshida+2003, OShea+2008}. 
The run yields results that are nearly identical to the fiducial case: one light-seed BH forms at $z\sim22$, followed by the formation of two massive seeds. 
The blue line in Extended Data Fig.~\ref{figex10:resolution} shows the evolution of the most massive BH in the higher-resolution run, demonstrating that the mass growth is largely insensitive to the numerical resolution achieved in our default simulation.

To examine uncertainties in the semi-analytic predictions, we also ran simulations with lower FUV background intensities. 
In our semi-analytic model, the stellar component of the source galaxy is populated using standard galaxy-formation prescriptions, but the star formation rate and efficiency are subject to substantial uncertainty. 
A lower star formation efficiency would result in a smaller luminosity and therefore a reduced FUV intensity incident on the massive BH forming halo. 
The reduced-FUV runs also capture possible spatial variations in the radiation field, which are not explicitly modeled in our semi-analytic treatment. 
If heavy seeds still form under weaker FUV irradiation, this supports both the robustness of the semi-analytic predictions and the insensitivity to spatial variations.

The green, yellow, and red lines in Extended Data Fig.~\ref{figex10:resolution} show the BH mass evolution when the background intensity is reduced by factors of two, five, and ten, respectively. 
The dependence on the seed formation epoch and subsequent growth is weak. 
As the FUV intensity decreases, the most massive BH forms slightly earlier, but the difference in formation redshift between the fiducial run and the lowest-intensity case is less than $\Delta z \simeq 0.1$. 
This behavior is consistent with the expectation that weaker FUV irradiation allows more efficient H$_2$ cooling, leading to earlier collapse of the target halo. 
The later mass growth is also only mildly affected: lower FUV intensity results in a somewhat smaller final BH mass, but the variation remains within a factor of three. 
Even in the weakest-FUV case, the final BH reaches $1.9\times10^7~M_\odot$ by $z=7$, highlighting the robustness of overmassive BH formation despite model uncertainties.

\bmhead{Data availability}
The simulation input files used in this study, including the runtime parameter file, the compile-time configuration file, and the initial conditions are available at https://doi.org/10.5281/zenodo.21242352. The full simulation outputs are available from the corresponding author upon reasonable request.

\bmhead{Code availability}
The simulation code used in this study is available as an archived source-code package at https://doi.org/10.5281/zenodo.21242352. Snapshot visualizations and the associated analyses were performed using particle\_visualization v1.0.0, which is publicly available on Zenodo (https://doi.org/10.5281/zenodo.21761582).

\bmhead{Acknowledgements} 
We thank Kohei Inayoshi for constructive discussions. We conduct numerical simulation on xd-2000 at the Center for Computational Astrophysics (CfCA) of the National Astronomical Observatory of Japan.

\bmhead{Funding Statement} 
This work was supported by JSPS KAKENHI Grant Numbers JP21H01122 (T.I), JP21K13960, JP21H01123, and JP26K00743 (S.H.). T.I has been supported by IAAR Research Support Program in Chiba University Japan, MEXT as ``Program for Promoting Researches on the Supercomputer Fugaku'' (JPMXP1020230406), and JICFuS. V.S. acknowledges support through a Leibniz Award by the Deutsche Forschungsgemeinschaft (DFG).

\bmhead{Author contributions} 
S.C. led the project, performed the hydrodynamical simulations, analysed the simulation data, produced the figures and wrote the manuscript. T.I. provided the initial conditions for black hole formation and contributed to the selection of the target haloes. S.H. contributed to the scientific motivation and interpretation of the study, including the connection between the semi-analytic model of T.I. and S.H. (2025) and the hydrodynamical simulations. V.S. contributed to the development and implementation of the simulation code, advised on the numerical methods and supervised the overall project. S.J.C. contributed to the radiation-transfer modelling and to the interpretation of the results in connection with observations. All authors discussed the results and commented on the manuscript.

\bmhead{Competing interests}
The authors declare no competing interests.

\clearpage

\begin{appendices}
\section*{Extended Data} 

\setcounter{figure}{0}
\renewcommand{\figurename}{Extended Data Fig.}
\renewcommand{\figureautorefname}{Extended Data Fig.}

\renewcommand{\tablename}{Extended Data Table.}

\begin{figure}[h]
\centering
\includegraphics[width=0.7\textwidth]{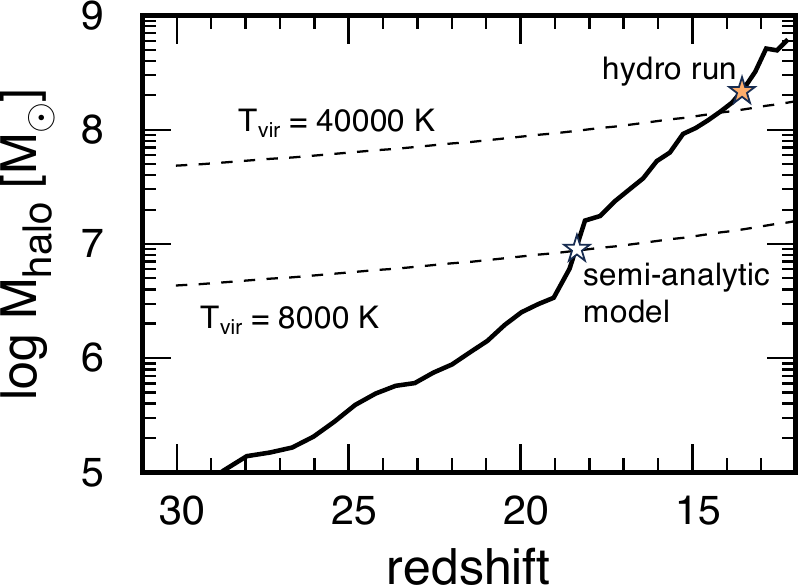}
\caption{
Redshift evolution of the halo mass in which MBH1 forms. The dashed lines indicate halo masses corresponding to virial temperatures of $8000$ and $40000$~K. The white star symbol marks the redshift at which heavy-seed formation is expected based on the semi-analytic model of \citet{Ishiyama+2025}, while the orange star symbol indicates the redshift at which heavy-seed formation occurs in our radiation-hydrodynamic simulation. The delayed onset of cloud collapse allows the halo to accumulate a larger gas reservoir, leading to the formation of an extremely massive seed BH.
}\label{figex1:Mhalo_evo}
\end{figure}

\begin{figure}[h]
\centering
\includegraphics[width=0.7\textwidth]{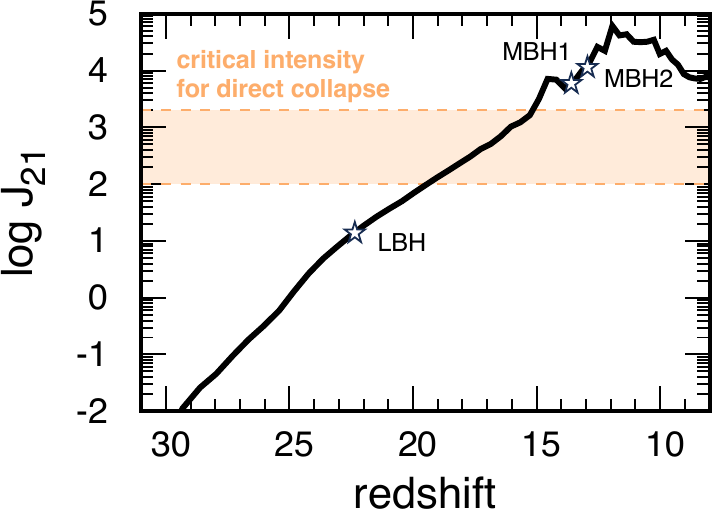}
\caption{
Redshift evolution of the LW radiation intensity ($J_{21}$) at the location of the target halo. The LW flux gradually increases over time as the stellar mass in the nearby source galaxy grows and the target halo moves closer to the radiation source. The orange shaded region denotes the critical intensity range required for the formation of massive BH seeds ($J_{21,\mathrm{crit}} \simeq 10$–$2000$) \citep{Shang+2010, Sugimura+2014, Latif+2014}. 
The redshifts at which the LBH, MBH1, and MBH2 seeds form are also shown.
}\label{figex2:J21_evo}
\end{figure}

\begin{figure}[h]
\centering
\includegraphics[width=0.7\textwidth]{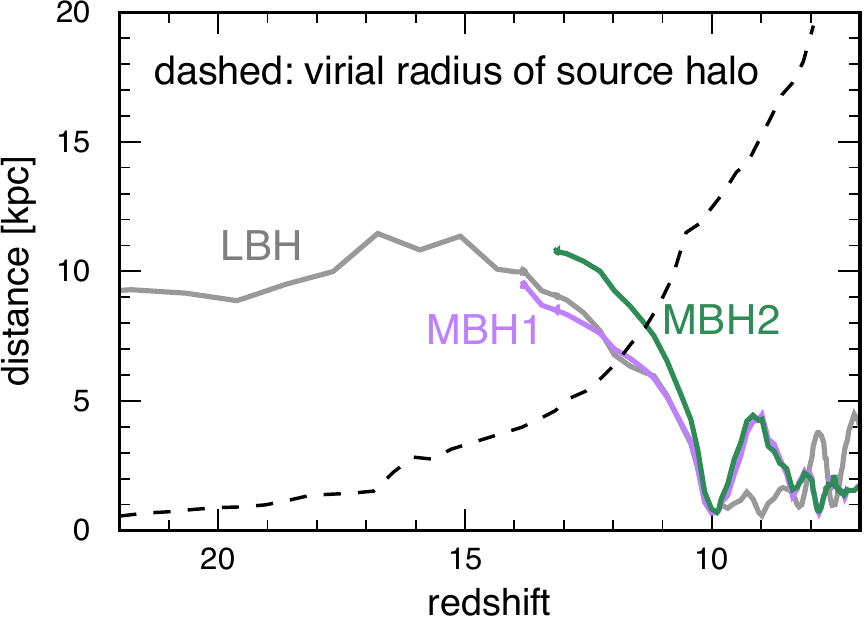}
\caption{
Time evolution of the distance between each seed BH and the source halo. The gray, purple, and green lines show the separations of the light-seed BH (LBH), the first massive BH (MBH1), and the second massive BH (MBH2), respectively. The dashed line indicates the evolution of the virial radius of the source halo. The seed BHs are captured by the source halo at redshifts $z\simeq12$ for LBH and MBH1, and $z\simeq11.2$ for MBH2.
}\label{figex7:migration}
\end{figure}

\begin{figure}[h]
\centering
\includegraphics[width=0.9\textwidth]{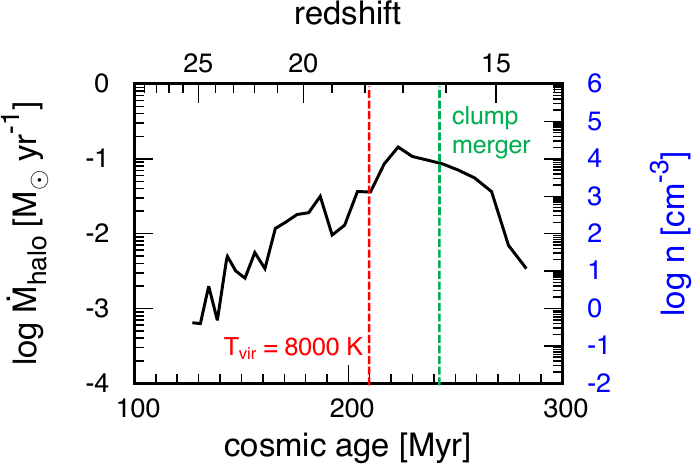}
\caption{
Time evolution of the halo mass growth rate (black) and the peak gas density (blue lines) in the host halo of MBH1. The dashed lines indicate two important epochs: the red line marks the time when the halo virial temperature reaches $8000~\mathrm{K}$ and the halo is identified as a DCBH halo, while the green line marks the merger of the clump in the DCBH halo with a nearby clump.
}\label{figex8:host_halo_evolution}
\end{figure}

\begin{figure}[h]
\centering
\includegraphics[width=1.\textwidth]{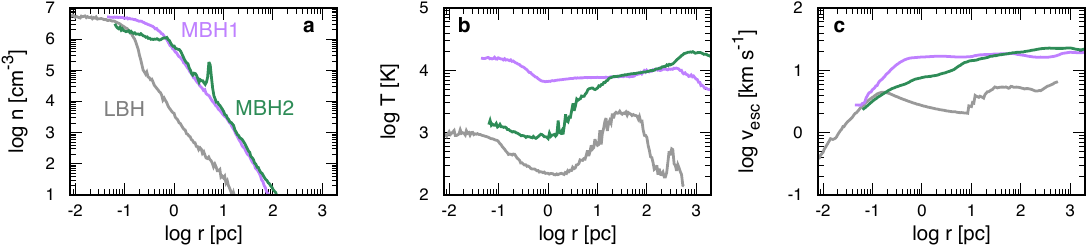}
\caption{
Radial profiles of gas density (a), temperature (b), and escape velocity (c) as functions of distance from the forming protostars.  The gray, purple, and green lines correspond to the epochs when the light-seed BH forms at $z\simeq22$ (LBH) and when the two massive-seed BHs form (MBH1 and MBH2), respectively.  
}\label{figex4:radial_profile}
\end{figure}

\begin{figure}[h]
\centering
\includegraphics[width=0.9\textwidth]{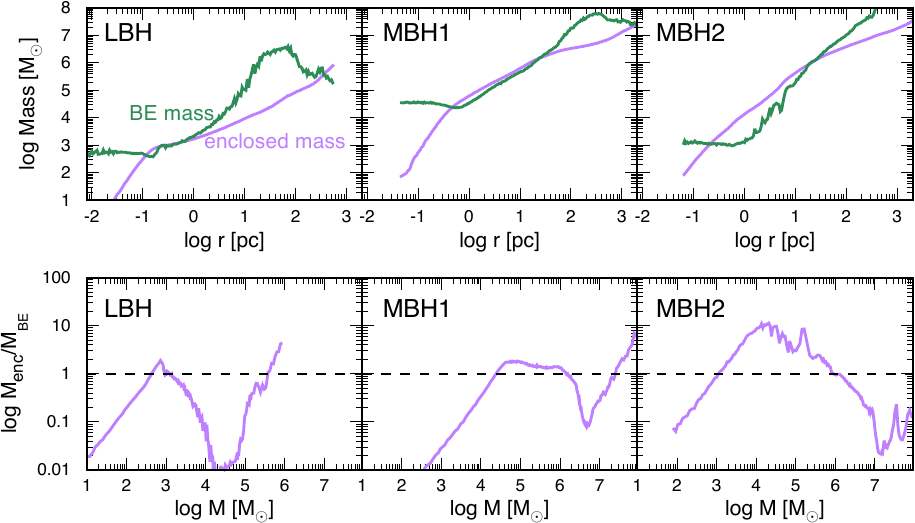}
\caption{
(top) Radial profiles of the Bonnor–Ebert (BE) mass (green) and the enclosed gas mass (purple).  Panels from left to right show the profiles for the light-seed (LBH) and massive-seed (MBH1 and MBH2) formation cases.  Regions where the enclosed mass exceeds the BE mass are gravitationally unstable and prone to collapse.  
(bottom) Radial profiles of the ratio between the enclosed mass and the BE mass ($M_\mathrm{enc}/M_\mathrm{BE}$).  The dashed line marks $M_\mathrm{enc}=M_\mathrm{BE}$, above which the cloud becomes unstable to gravitational collapse.  Only the innermost $\sim10^3\,M_\odot$ region becomes unstable in the LBH case,  whereas up to $\sim10^6\,M_\odot$ of gas is unstable in the MBH1 and MBH2 cases.
}\label{figex5:Menc_MBH}
\end{figure}

\begin{figure}[h]
\centering
\includegraphics[width=0.9\textwidth]{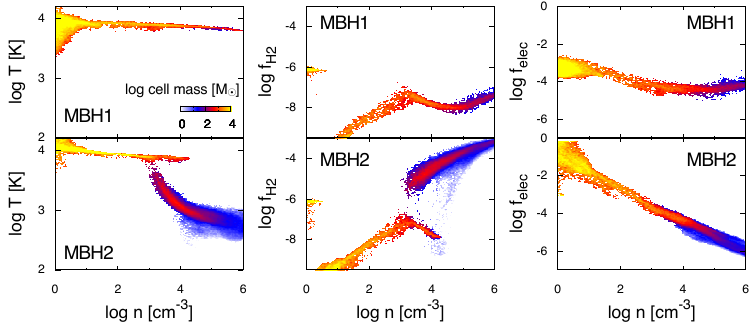}
\caption{
Two-dimensional histograms of temperature (left), molecular hydrogen fraction (middle), and electron fraction (right) as functions of gas density. The top and bottom panels show the results for MBH1 and MBH2, respectively. Each panel is divided into $200\times200$ bins, with colors indicating the gas mass in each bin.
}\label{figex6:2dhist}
\end{figure}

\begin{figure}[h]
\centering
\includegraphics[width=0.9\textwidth]{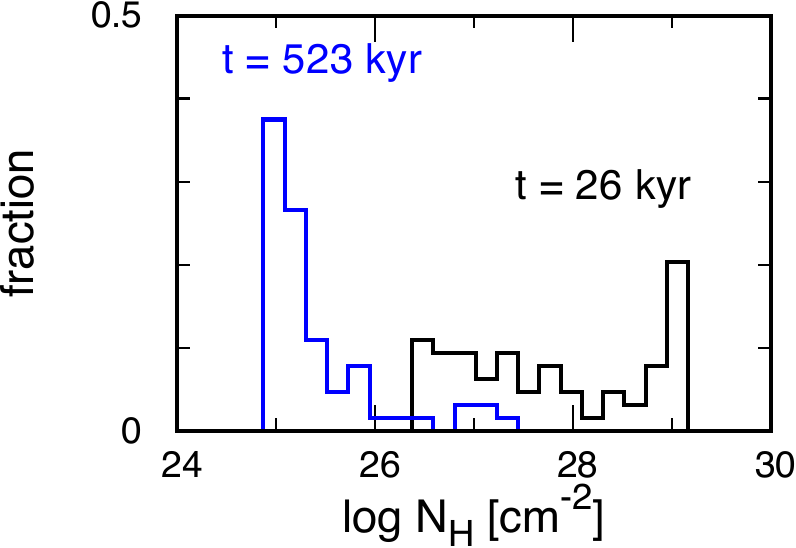}
\caption{
Distribution of the column density around MBH2 at two different epochs, $t=26~\mathrm{kyr}$ (black) and $512~\mathrm{kyr}$ (blue). We cast rays uniformly along 64 different directions and measure the column density along each ray.
}\label{figex3:NH_distribution}
\end{figure}

\begin{figure}[h]
\centering
\includegraphics[width=1.0\textwidth]{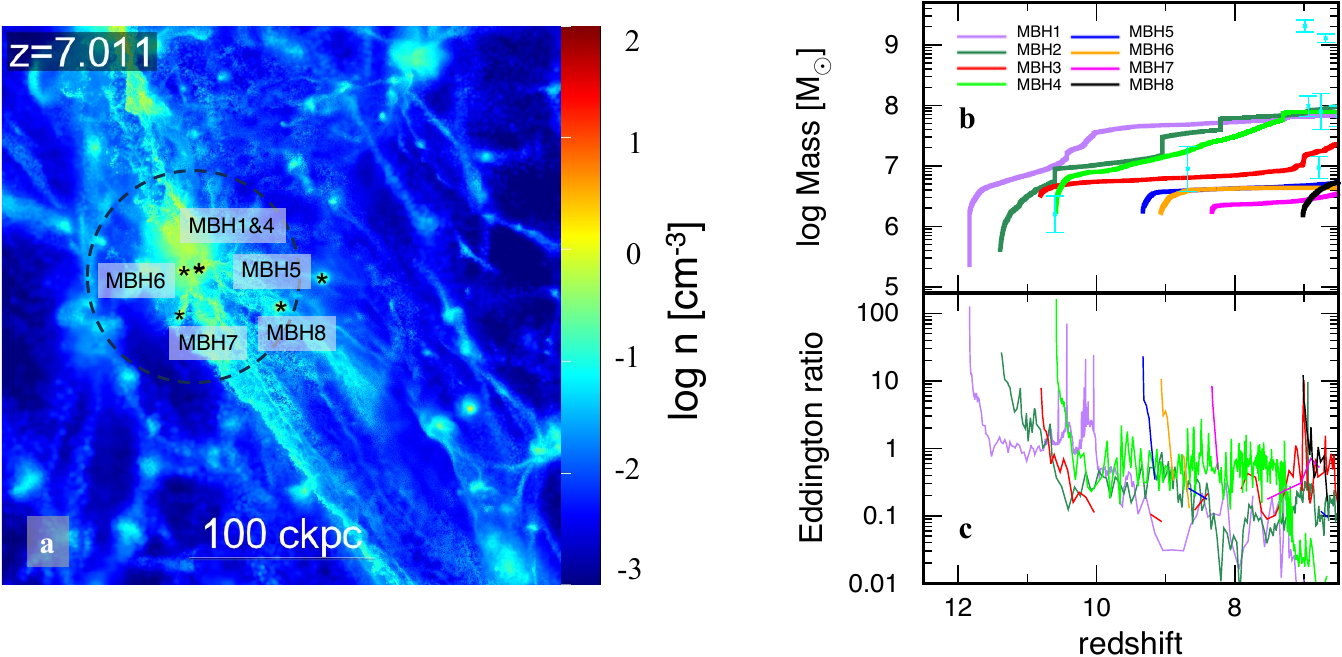}
\caption{
Spatial distribution of BHs in the zoom-in simulation of the later DCBH formation scenario. By $z=7.011$, eight BHs have already formed; in panel a, the color scale indicates the gas density. The dashed circle shows the virial radius of the source halo. Note that MBH2 and MBH3 is located outside this panel. Panels b and c show the redshift evolution of the BH mass and Eddington ratio, respectively, for the eight BHs formed in this region. The cyan symbols indicate the mass and redshift distribution of observed high-redshift AGNs and QSOs.
}\label{figex9:MBH_evolution_ID22}
\end{figure}

\begin{figure}[h]
\centering
\includegraphics[width=0.7\textwidth]{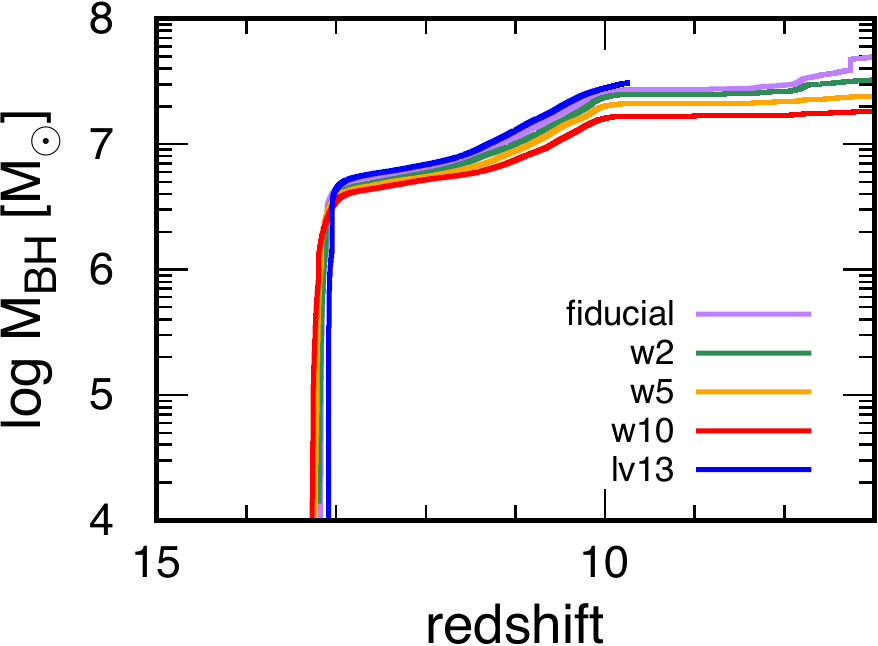}
\caption{
Time evolution of the most massive BH in the comparison runs. 
The purple line shows the fiducial run, referred to as MBH2 in the main text. The green, yellow and red lines show the evolution when the luminosity of the source halo in the LW band is reduced by factors of two, five and ten, respectively. The blue line shows the evolution in the higher-resolution run.
}\label{figex10:resolution}
\end{figure}

\end{appendices}

\clearpage

\end{document}